# Azimuthal Instability in Oscillating Liquid Jets by the Weber Number Variations

Hyoungwoo Kim (김형우),[1] Seungwon Shin (신승원),[2] Hyosang Yoon (윤효상),[1] Sanghyuen Park (박상현),[1] and Dong-hyuk Shin (신동혁)[1,a)]

*[1]Department of Aerospace Engineering, Korea Advanced Institute of Science and Technology, Daejeon 34141, South Korea*

*[2]Department of Mechanical and System Design Engineering, Hongik University, Seoul 04066, South Korea*

This paper investigates the dependence of azimuthal mode growth rate of oscillating liquid laminar jets on the Weber number. The oscillating liquid laminar jets were simulated over a range of Weber numbers. The simulation setup was validated by comparing the instantaneous interface shapes and the visible azimuthal mode from experimental measurements. As the Weber number increases, the spreading angle rises due to the small surface tension, and higher azimuthal modes are observed. Fourier Transform analysis is used to find the exact azimuthal mode of the simulation results. The dominant azimuthal mode increases as the Weber number increases. A power-law dependence of the growth rate on the Weber number is identified, with the obtained exponents showing good agreement with linear instability analysis. Furthermore, the rescaled Fast Fourier Transform (FFT) results collapse onto a single growth curve for each azimuthal mode, validating the proposed power-law for predicting the growth rates of oscillating liquid jets.

a) Authors to whom correspondence should be addressed: donghyuk.shin@kaist.ac.kr

## I. INTRODUCTION

Periodic oscillations in the flow field are widely observed in many configurations. Vortex shedding behind a cylinder and pressure oscillations in a closed chamber are such examples. In combustion chambers, the unsteady heat release can be coupled with pressure oscillations so that an undesirable phenomenon, the so-called combustion instability, can happen.[1] If combustion instability happens, the temperature and the pressure locally increase than the design point, so the instability can lead to failure of the engine.

Combustion instability is a major technological challenge in the development of many propulsion systems, such as gas turbines and liquid rocket engines. In an extreme case, the combustor may explode, leading to cost and safety concerns and the potential to severely delay development.[2] Although many previous studies have sought to understand the full mechanisms of combustion instability[3-7], the detailed relationship induced by pressure fluctuations remains difficult to predict and control.

In most such engines, spray combustion is frequently used. In spray combustion, liquid reactants undergo atomization, vaporization, and chemical reactions. Among these processes, spray characteristics, such as droplet size distribution, spatial distribution, and liquid surface area, are strongly determined by the atomization process and have been shown to be a critical factor influencing combustion instability.[8-11] Although spray dynamics and breakup have long been studied,[12-16] the fundamental physics of fuel sprays is still regarded as insufficiently understood. Numerical investigations of turbulent fuel jet flames have also emphasized the importance of flow development and fuel mixing in determining the local reactive state.[17] Lieuwen and Zinn[5] suggested a specific mechanism of combustion instability that is induced by coupled interactions between pressure perturbations at the flame and fuel supply rates in gas turbines. A starting point for further exploration of their proposed mechanism is to identify that a key component must include the physics of pressure-induced oscillations in liquid jets.

In general, three primary instability mechanisms are typically attributed to the jet breakup: (*i*) the Rayleigh-Taylor instability (RTI), arising from the density difference between two fluids; (*ii*) the Rayleigh-Plateau instability (RPI), driven by surface tension; and (*iii*) the Kelvin-Helmholtz instability (KHI), induced by the shear stress between the liquid jet and the ambient gas. Krechetnikov and Homsy[18] explained that the crown spike formation is induced by the Richtmyer-Meshkov instability, which can be regarded as an impulsive-acceleration case of the RTI[19]. When jet atomization occurs in crossflow, the KHI was shown to induce column bag breakup[20]. Based on linear instability analysis, Qiao et al.[21] showed that in an oscillating liquid jet, a moderate longitudinal wavenumber is induced by the combination of the RPI and KHI (RP-KHI), and short longitudinal wavenumbers were driven by the oscillating Kelvin-Helmholtz instability (OKHI).

Liquid jet instabilities have been extensively investigated, with particular attention given to the development of disturbances along the jet axis. Keller et al.[22] established a spatial instability framework for liquid jets by treating the wavenumber as complex and the frequency as real, thereby describing the spatial amplification of disturbances along the jet. Busker et al.[23] subsequently investigated the nonlinear breakup of an inviscid liquid jet and examined the dependence of the break-off mode on the disturbance amplitude and wavelength. Chauhan et al.[24] experimentally investigated the downstream evolution of disturbances in a liquid jet and characterized their wavelength and growth rate. These studies have predominantly focused on the longitudinal development and breakup of liquid jets. Non-axisymmetric instabilities have also been investigated, although to a considerably lesser extent. Yang[14], for example, analyzed the asymmetric instability of a liquid jet and demonstrated the development of unstable non-axisymmetric disturbances under appropriate flow conditions. Nevertheless, compared with the extensive studies of longitudinal instability, the characteristics and mode selection of azimuthal instability have received comparatively limited attention. This trend is also observed in oscillating liquid jets, where externally imposed flow or pressure oscillations can further promote jet instability and breakup.

Oscillating liquid jets have therefore been investigated extensively to understand the effects of periodic forcing on jet breakup. Dressler[25] found that large-amplitude velocity perturbations drive instabilities on a cylindrical liquid jet, breaking it into droplets. Chaves et al.[26] and Geschner et al.[27] studied the disintegration of a sinusoidally forced liquid jet and classified jet morphologies and spray structures under varying modulation characteristics, injection pressure, and jet velocity. Numerical studies have also been conducted to investigate the effects of periodic forcing on the breakup dynamics. Srinivasan et al.[28] conducted numerical simulations based on the volume of fluid (VOF) method to characterize the breakup process under varying modulation amplitude while maintaining constant Strouhal numbers. Zhang and Shin[29] investigated the effects of ambient pressure oscillation on cylindrical liquid jet sprays, using the VOF method. They found that the pressure amplitude oscillation has the most significant impact on the primary breakup, and the surface tension highly influences the shape and number of droplets once ligaments form. Although these studies have provided substantial insight into the instability and breakup dynamics of oscillating liquid jets, an additional azimuthal instability has also been observed.[25, 27] In experiments,[25, 27] a significant azimuthal instability of the oscillating liquid jet manifests after only 3 to 5 forcing periods. However, the development and mode characteristics of this azimuthal instability remain insufficiently understood. Thus, in this work, a three-dimensional simulation is conducted to capture the azimuthal instability.

The azimuthal instability is also found in vortex rings. In the head-on collision experiment of two vortex rings, the development of azimuthal waves and the eventual formation of small rings are observed[30]. Cheng et al.[31] reproduced the azimuthal waves with numerical simulation. The azimuthal instability of a liquid can also be found in liquid droplet impact. Rioboo et al.[32] identified the transition boundaries between deposition, crown formation, and crown splash in terms of the Weber number. As the Weber number increases, the impact regime changes from deposition to crown formation and eventually to crown splash. Since higher Weber numbers promote the development of azimuthal instabilities, in this study, numerical simulations are performed over a range of Weber numbers.

In this study, the azimuthal instability of the oscillating liquid jet is investigated with numerical simulations. In the first step, the oscillating liquid jet simulation is conducted with the same dimensionless numbers as the experimental result[27] for validation. Then, the numerical simulations are performed by varying the Weber number. Moreover, the correlation between the Weber number and the growth of azimuthal instability is investigated.

## II. NUMERICAL METHODS

In this study, a coupled solver[33] of PeleLM[34-36] and Interface Reconstruction Library (IRL)[37-39] is employed. PeleLM solves the variable-density, reacting Navier-Stokes equations in the low Mach limit within an adaptive mesh refinement framework. PeleLM is selected for subsequent application to coupled combustion problems. IRL provides routines to implement a geometric VOF scheme to compute the flux of the volume of fluid using the Efficient Least-squares VOF Interface Reconstruction Algorithm (ELVIRA) 3D[40] method. A coupled solver, PeleLM+IRL[33], solves the multiphase flows including the surface tension. The governing equations of PeleLM+IRL are:

$$\alpha_t + \nabla \cdot (\mathbf{U}\alpha) = 0. \tag{1}$$

$$\mathbf{U}_t + (\mathbf{U} \cdot \nabla)\mathbf{U} = \frac{1}{\rho}(-\nabla p + \mu \nabla^2 \mathbf{U} + \vec{F}_V), \tag{2}$$

For nonreacting isothermal flows,

$$\nabla \cdot \mathbf{U} = 0\,, \tag{3}$$

where $\alpha$, $\boldsymbol{U}$, $\rho$, and $p$ represent the volume fraction, velocity, density, and pressure, respectively. $\vec{F}_V$ represents external volume forces. Here, $\mu$ is the dynamic viscosity coefficient. The subscript $t$ refers to the time derivative. A surface tension force is incorporated into the original momentum equation (Eq. (2)) as an external volume force term, $\vec{F}_V$.

This solver was validated with (*i*) translations and rotations of Zalesak's disk, (*ii*) a stationary circular droplet with surface tension, (*iii*) an oscillating elliptical droplet, and (*iv*) three-dimensional deformation of a spherical droplet[33]. For further details, see Kim et al.[33] and the references therein.

## III. SIMULATION SETUP AND VALIDATION SIMULATION

Fig. 1(a) presents the oscillating liquid jet experiment[27]. In the experiment, the liquid was excited by a piezo-ceramic oscillator and fed into the resonance space. The pressure fluctuation within the resonance space can be intensified by a movable piston. The mushroom-shaped oscillation was captured by a CCD camera.

To mimic the experiments, a cuboid computational domain is designed as shown in Fig. 1(b). On the $x = 0$ plane, the liquid jet is injected into the computational domain. The liquid injector diameter is 0.6 mm, and the domain sizes in the *y*- and *z*- directions are both 6 mm, which is 10 times the diameter. The adaptive mesh refinement (AMR) is used with level 2, resulting in a locally refined mesh near the liquid jet, as shown in Fig. 1(b). A no-slip wall condition is used on the $x = 0$ plane except for the inlet. An outflow boundary condition, with zero normal gradients on all conserved variables, is utilized for the other boundaries.

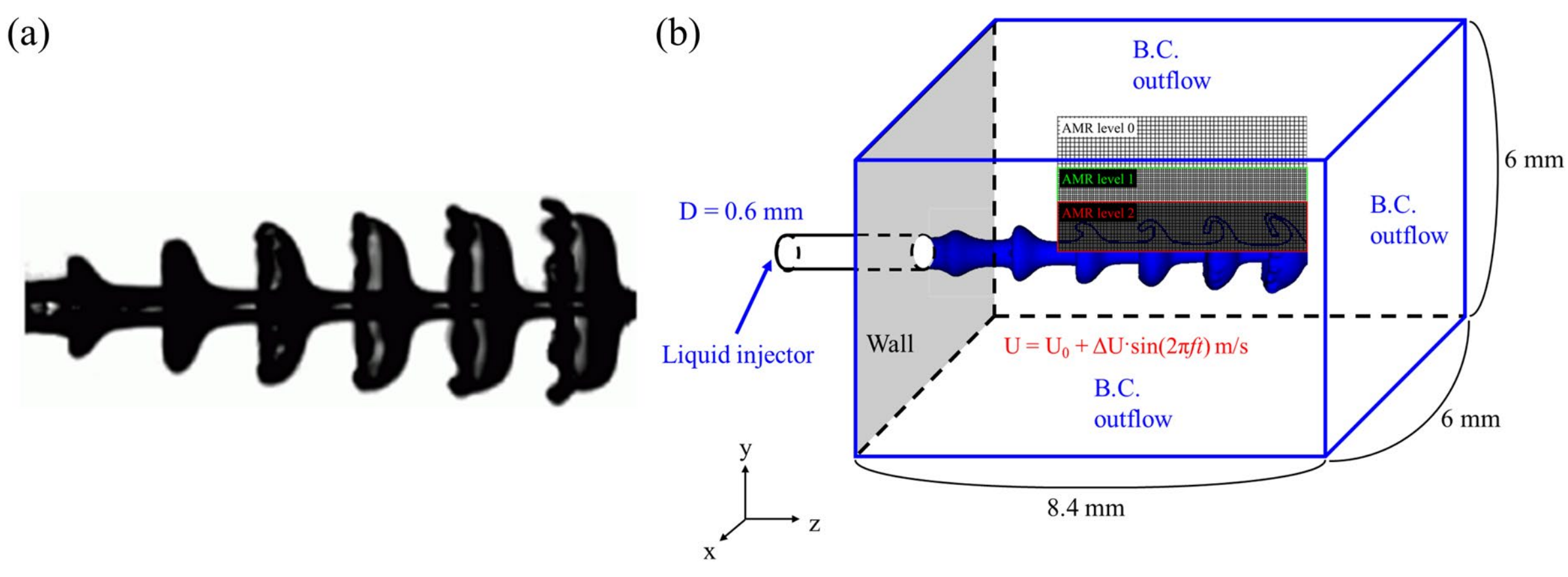

**Fig. 1. (a) Oscillating liquid jet[27] and (b) a schematic of the computational domain.**

The configuration is specified with four dimensionless numbers, Reynolds number (Re), Weber number (We), Strouhal number (St), and the magnitude ratio of velocity oscillation (ΔU/U):

$$\mathrm{Re} = \frac{\rho_l U D}{\mu_l};\ \mathrm{We} = \frac{\rho_l U^2 D}{\sigma};\ \mathrm{St} = \frac{\pi D f}{U};\ \frac{\Delta U}{U}. \tag{4}$$

In previous studies[18,32], the azimuthal instability changes with the Weber number. Thus, this work focuses on the variation of azimuthal instability with the Weber number. Several oscillating liquid jet simulations are conducted with various Weber numbers. Changing the surface tension coefficient is a simple method to change the Weber number. Thus, the surface tension coefficient is varied from 0.25 to 6 mN/m. The velocity and viscosity are set to 6 m/s and 0.0011 kg/m·s, respectively. The computational time decreases with decreasing velocity oscillation amplitude. Since azimuthal instability is observed even at $\Delta U/U = 0.1$, this value is adopted to reduce the computational cost. The simulation conditions are summarized in Table I.

**Table I. Simulation conditions**

| Re | We | St | D [mm] | U [m/s] | $f$ [Hz] | $\mu$ [kg/m·s] | $\sigma$ [mN/m] | ΔU/U |
|---|---|---|---|---|---|---|---|---|
| 3273 | 3600 ~ 86400 | 1.57 | 0.6 | 6 | 5000 | 0.0011 | 0.25 ~ 6 | 0.1 |

Regarding the grid resolution, three different grid resolutions are tested: (*i*) coarse, (*ii*) medium, and (*iii*) fine grids, whose grid sizes near the liquid jet are 0.0375 mm, 0.01875 mm, and 0.0125 mm, respectively. The grid study is conducted to determine an appropriate grid resolution for the azimuthal instability analysis of the mushroom-shaped oscillation. Among the simulation conditions listed in Table I for the azimuthal instability analysis, the Weber number of 86400 is selected for the grid study because it exhibits the highest azimuthal mode among the conditions considered. The coarse grid is 2 times coarser than the medium grid, and the fine grid is 1.5 times finer than the medium grid. Figure 2 presents the mushroom-shaped oscillation using the three grids at 1.4 msec. The azimuthal instability is not clearly resolved on the coarse grid, whereas it is apparent on the medium and fine grids.

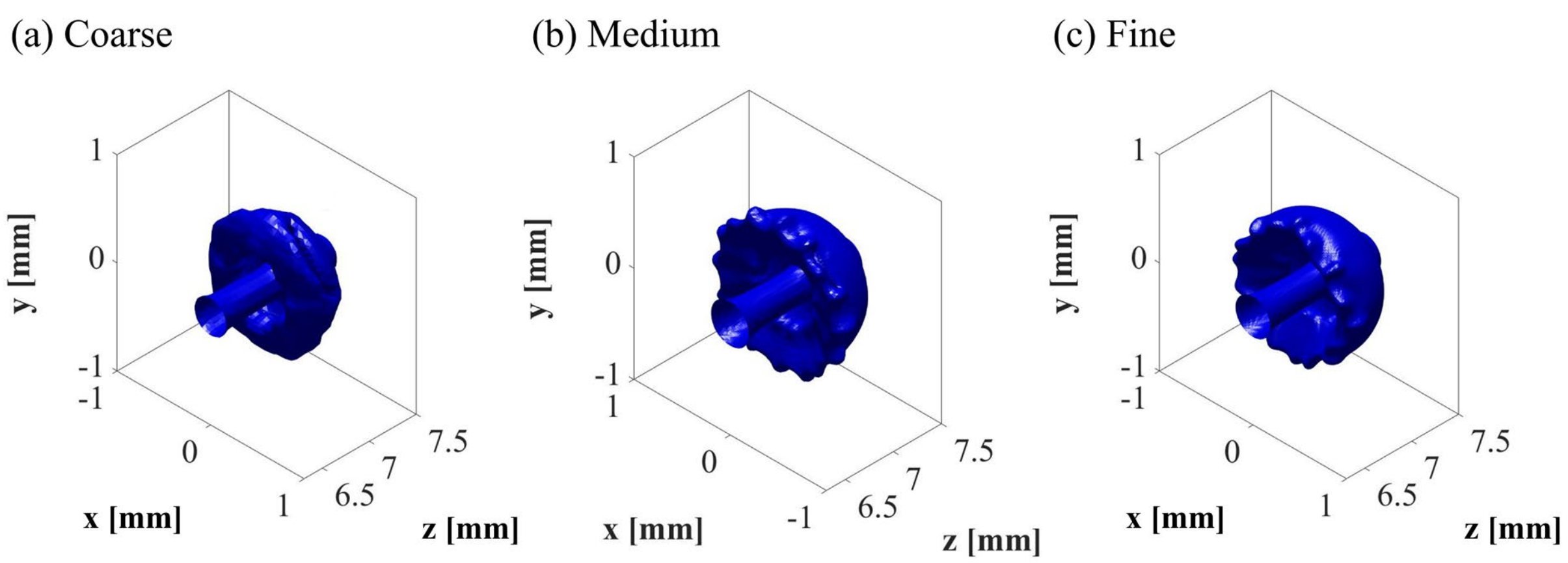

**Fig. 2. Mushroom-shaped oscillation using (a) coarse, (b) medium, and (c) fine grids at 1.4 msec.**

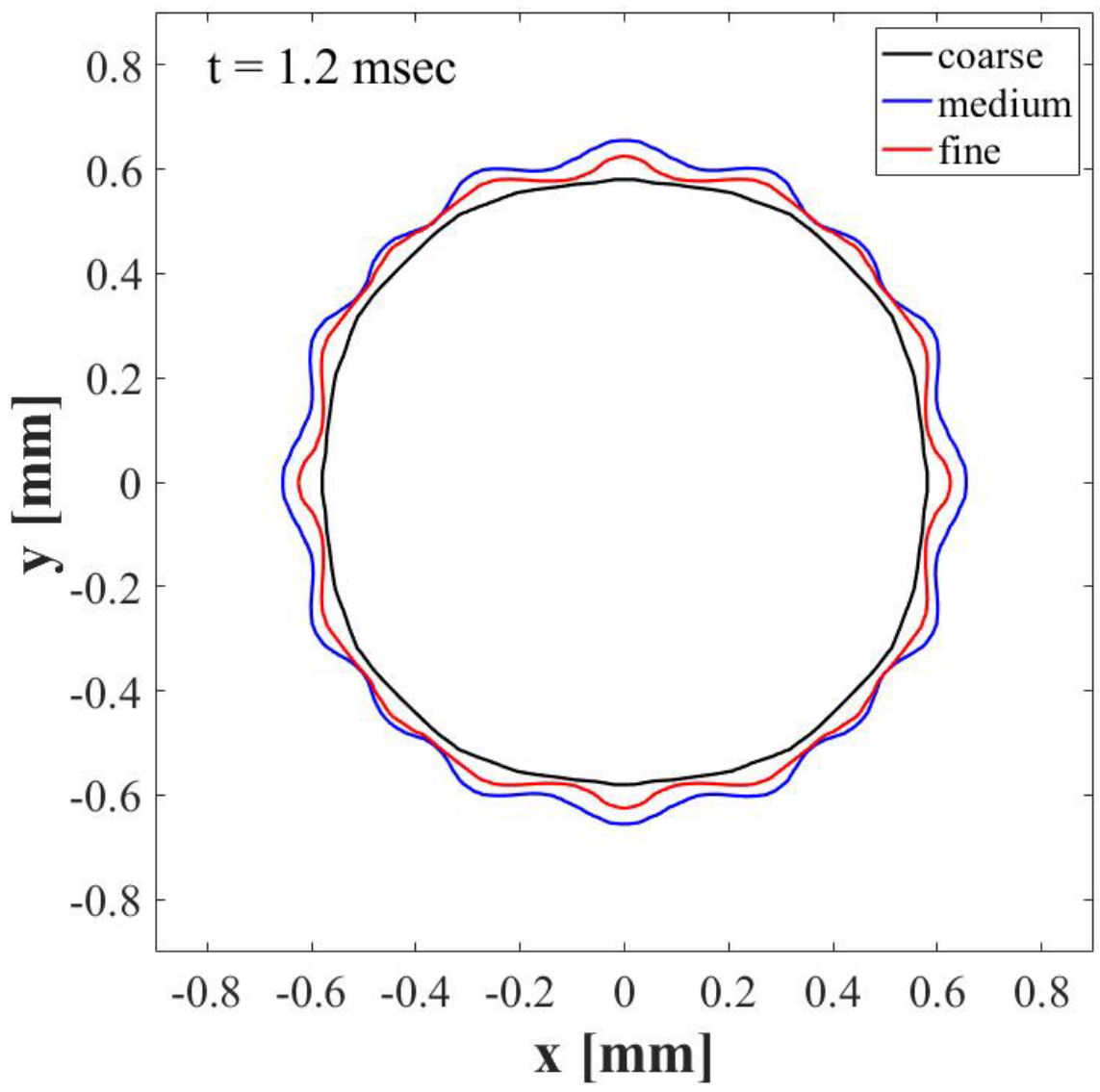

**Fig. 3. Outermost contour of the oscillating jet on the coarse, medium, and fine grids at 1.2 msec.**

Figure 3 depicts the outermost contour of the jet on the three grids at 1.2 msec. As shown in Fig. 2, the contour shape of the medium grid is similar to that of the fine grid, whereas the coarse grid's contour shape differs significantly. As time advanced, the azimuthal instability developed on the medium and fine grids, but not on the coarse grid.

Fast Fourier Transform (FFT) analysis is used to find the exact azimuthal mode. Figure 4 presents the processes to conduct the FFT analysis. First, Fig. 4(a) shows the contour of the oscillating liquid jet where the Weber number is 21600. Then, the *x*-*y* plot of the contour is converted to an $R$-$\theta$ plot (Fig. 4(b)). The $R$-$\theta$ data is used to conduct the FFT analysis using the following equation.

$$R = a_0 + \sum_{m=1}^{\infty} A_m \cos(2\pi m\theta) \tag{5}$$

The $m$ means the azimuthal mode. Figure 4(c) presents the result of the FFT analysis for a Weber number of 21600. The magnitude means $|A_m|$. In this case, modes 4 and 12 are shown, and mode 12 is dominant.

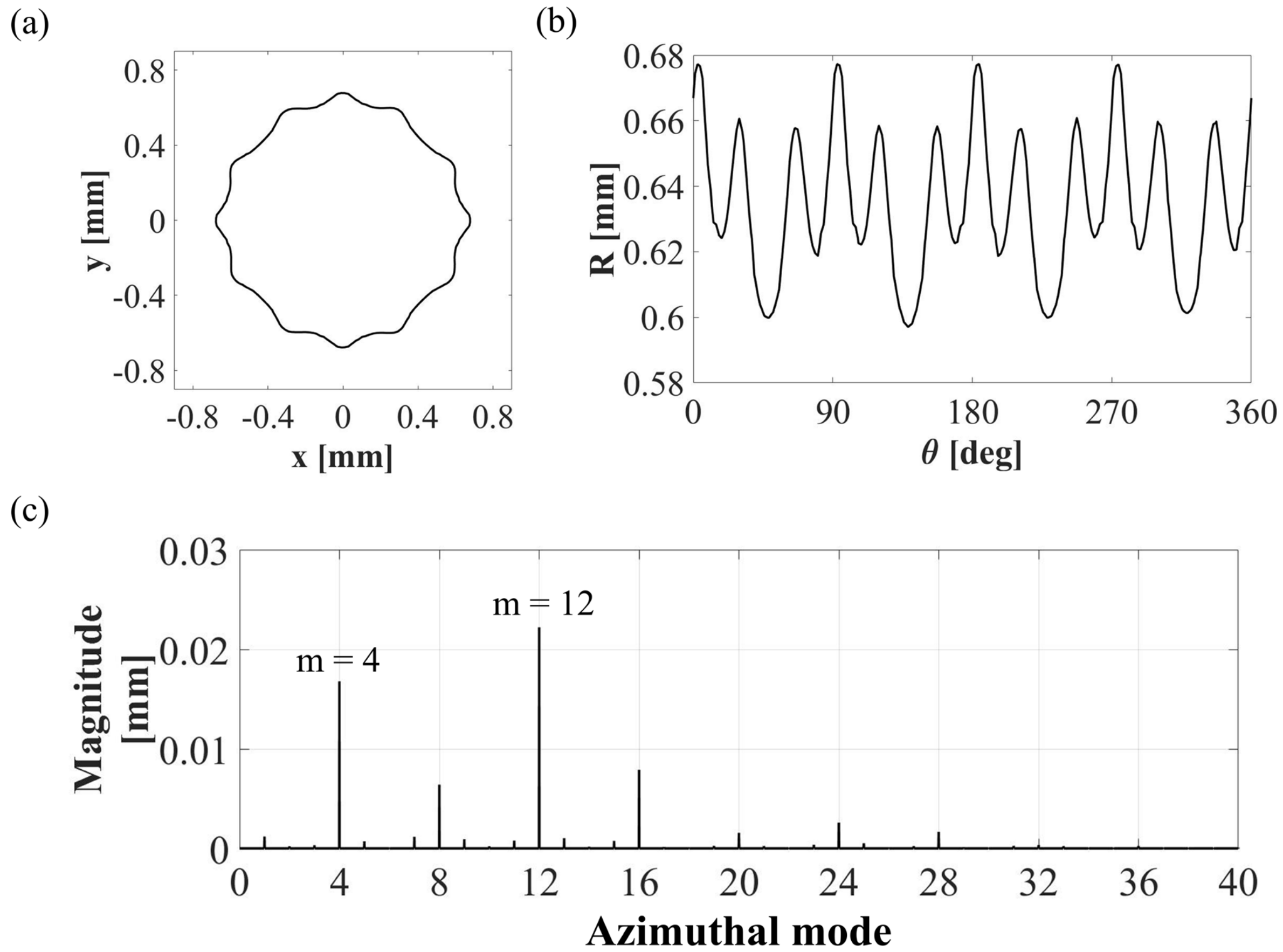

**Fig. 4. (a) Contour of the oscillating liquid jet where We = 21600, (b) an *R*-*θ* plot of the contour, and (c) the FFT result of the *R*-*θ* plot.**

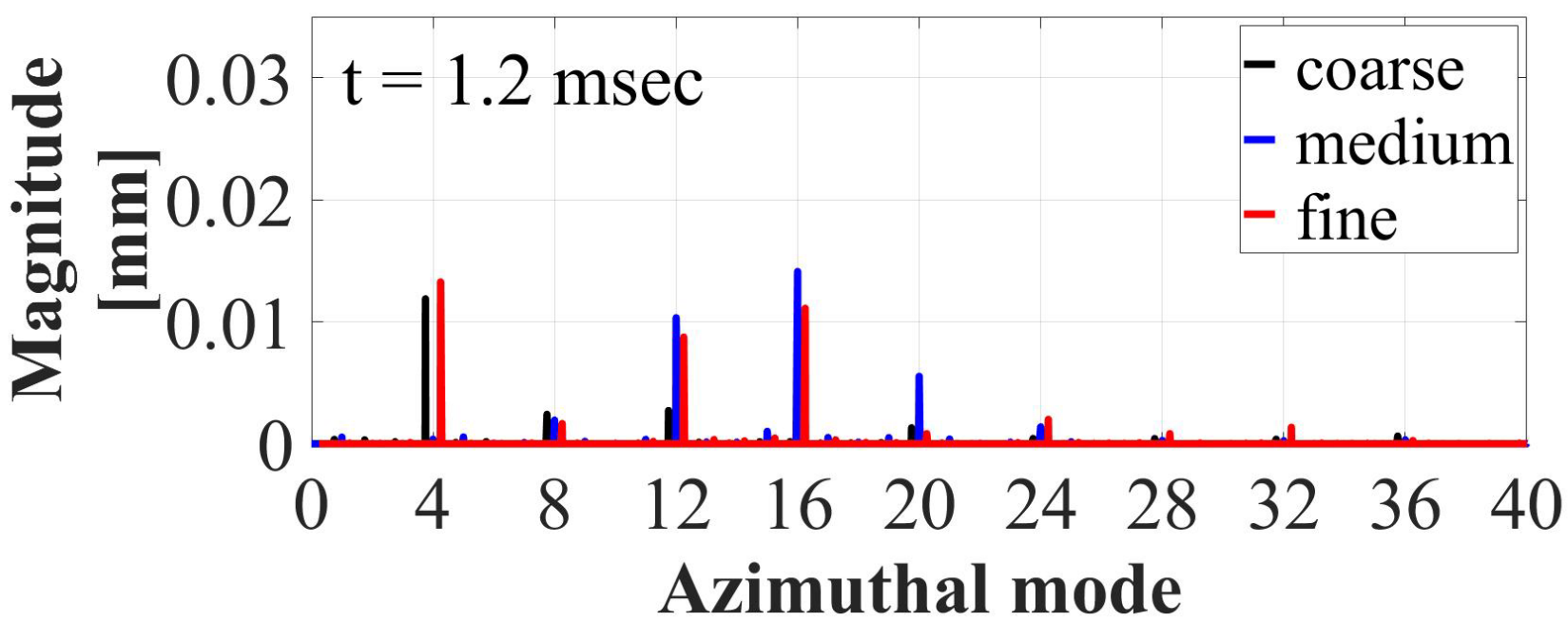

**Fig. 5. The Fast Fourier Transform (FFT) results of the coarse, medium, and fine grids at 1.2 msec.**

The Fast Fourier Transform (FFT) results of the coarse, medium, and fine grids are shown in Fig. 5. In both the medium and fine grids, magnitudes of 12 and 16 modes appear, while magnitudes of 12 and 16 modes in the coarse grid are small. Although the fine grid shows a relatively stronger low-order mode at m = 4, the primary spectral content and dominant modes at higher wavenumbers are consistently captured on the medium grid. This indicates that the medium grid adequately resolves the main instability features. Thus, the medium grid is adopted for subsequent simulations, considering computational efficiency.

Using the medium grid, the simulation configuration is validated by comparing the experimental result[27]. The same dimensionless numbers that characterize the experimental configuration[27] are used to validate the oscillating liquid jet simulation. The Reynolds number, the Weber number, and the Strouhal number are 3320, 3582, and 1.6, respectively. The experiments did not specify the magnitude of velocity oscillation. After several trials on ΔU/U, ΔU/U is chosen to be 0.225 because this value provides qualitatively comparable outcomes. The liquid and gas phases were assigned the same dynamic viscosity, as incorporating variable viscosity posed considerable numerical difficulties and adversely affected the robustness of the simulations. This treatment was adopted to facilitate stable and reliable computations. The mentioned validation simulation conditions are shown in Table II.

**Table II. Conditions for the validation simulation**

| Re | We | St | D [mm] | U [m/s] | $f$ [Hz] | $\mu$ [kg/m·s] | $\sigma$ [mN/m] | ΔU/U |
|---|---|---|---|---|---|---|---|---|
| 3320 | 3582 | 1.6 | 0.6 | 5.89 | 5000 | 0.001065 | 5.811 | 0.225 |

Figure 6 presents the oscillating liquid jet results of Geschner et al.[27] and the PeleLM+IRL simulation. The mushroom-shaped oscillations appear in a similar position, indicating that the liquid jet oscillates in a similar position. As the liquid jet develops, the thickness of the liquid column decreases, and the size of the mushroom-shaped oscillation increases. The azimuthal wave appears in the same oscillation, which is the fourth oscillation.

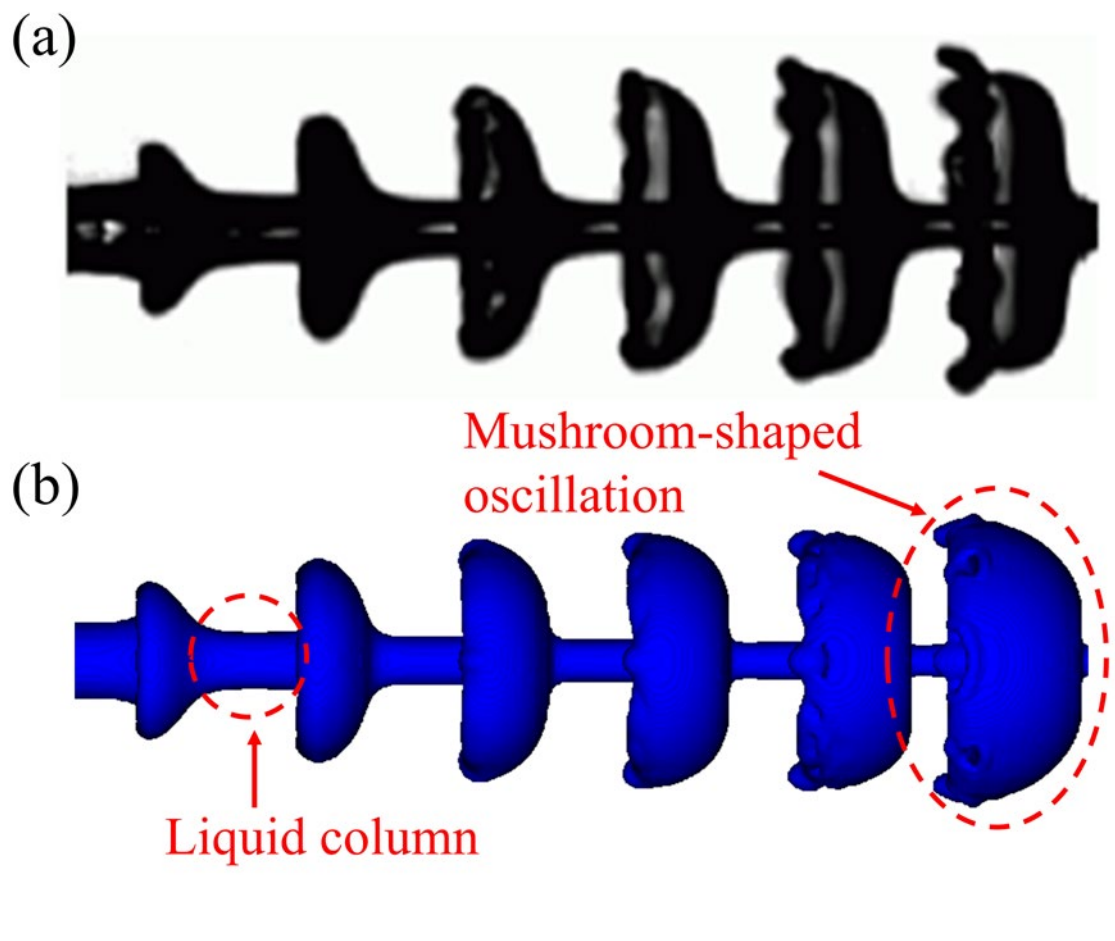

**Fig. 6. Oscillating liquid jet results: (a) Geschner et al.,[27] (b) PeleLM+IRL simulation.**

Figure 7 shows the comparison of oscillating liquid jet results. The dimensions of each result are normalized with the length of the liquid jet. Then, the contour data points of Geschner et al.[27] are extracted and plotted as the red dots in Fig. 7(a). The extracted red dots in Fig. 7(a) are shown with the contour of the PeleLM+IRL simulation in Fig. 7(b). The blue line is a contour from the PeleLM+IRL simulation. The normalized spread radius and the length of the mushroom-shaped oscillations are similar. The maximum radial discrepancy between experimental and simulation results is 26.9%. The maximum longitudinal discrepancy between experimental and Zhang and Shin's simulation results[29] is 29.0%, indicating that the present simulation achieves a comparable level of agreement with the experimental results.

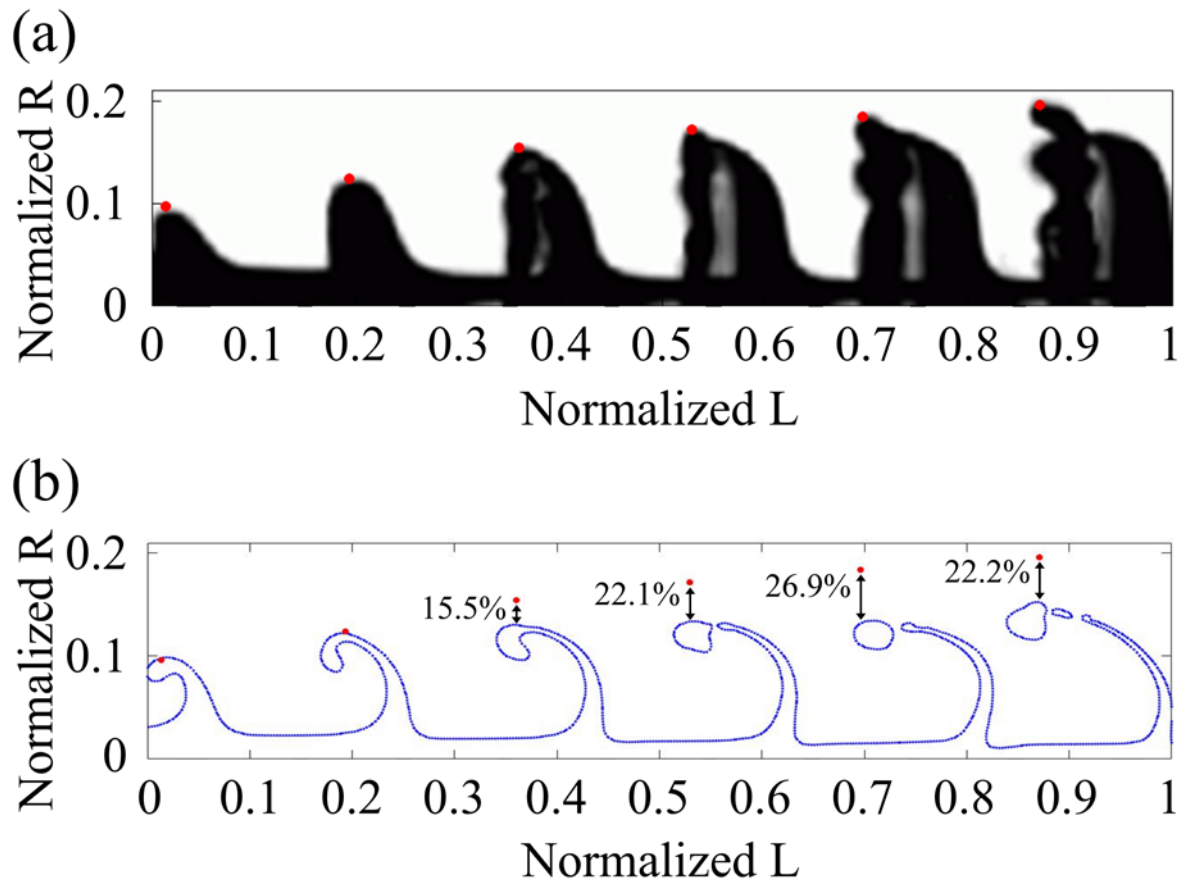

**Fig. 7. Comparison of oscillating liquid jet results: (a) dots on Geschner et al.,[27] (b) PeleLM+IRL simulation contour with the same dots.**

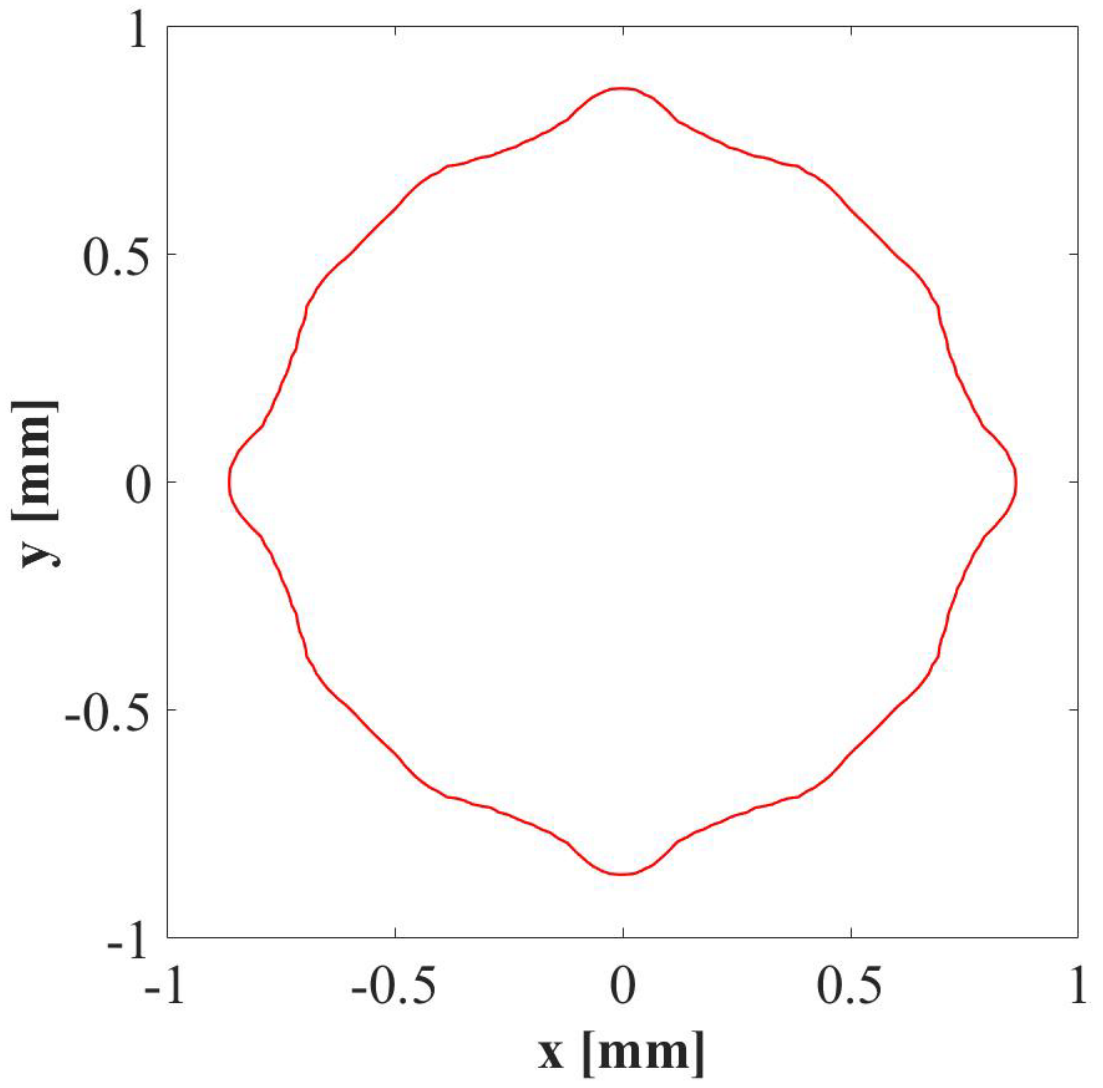

**Fig. 8. Mushroom-shaped oscillation contour of an oscillating liquid jet simulated using PeleLM+IRL.**

Figure 8 depicts a mushroom-shaped oscillation contour of an oscillating liquid jet simulated using PeleLM+IRL. The contour shows the outermost contour of a mushroom-shaped oscillation. The azimuthal instability is apparent in the contour, exhibiting 12 peaks. The azimuthal mode of an oscillating liquid jet simulation is 12. 8-12 peaks are observed in experimental results (Fig. 6(a)), indicating that the simulation reproduces azimuthal instability similar to that observed experimentally. The good agreement between the present simulations and experimental results suggests that the viscosity difference between the liquid and gas phases has a limited influence on the azimuthal instability under the conditions considered here. Thus, our simulation setup is suitable for analyzing azimuthal instability of oscillating liquid jets.

## IV. RESULTS

### A. Geometrical results

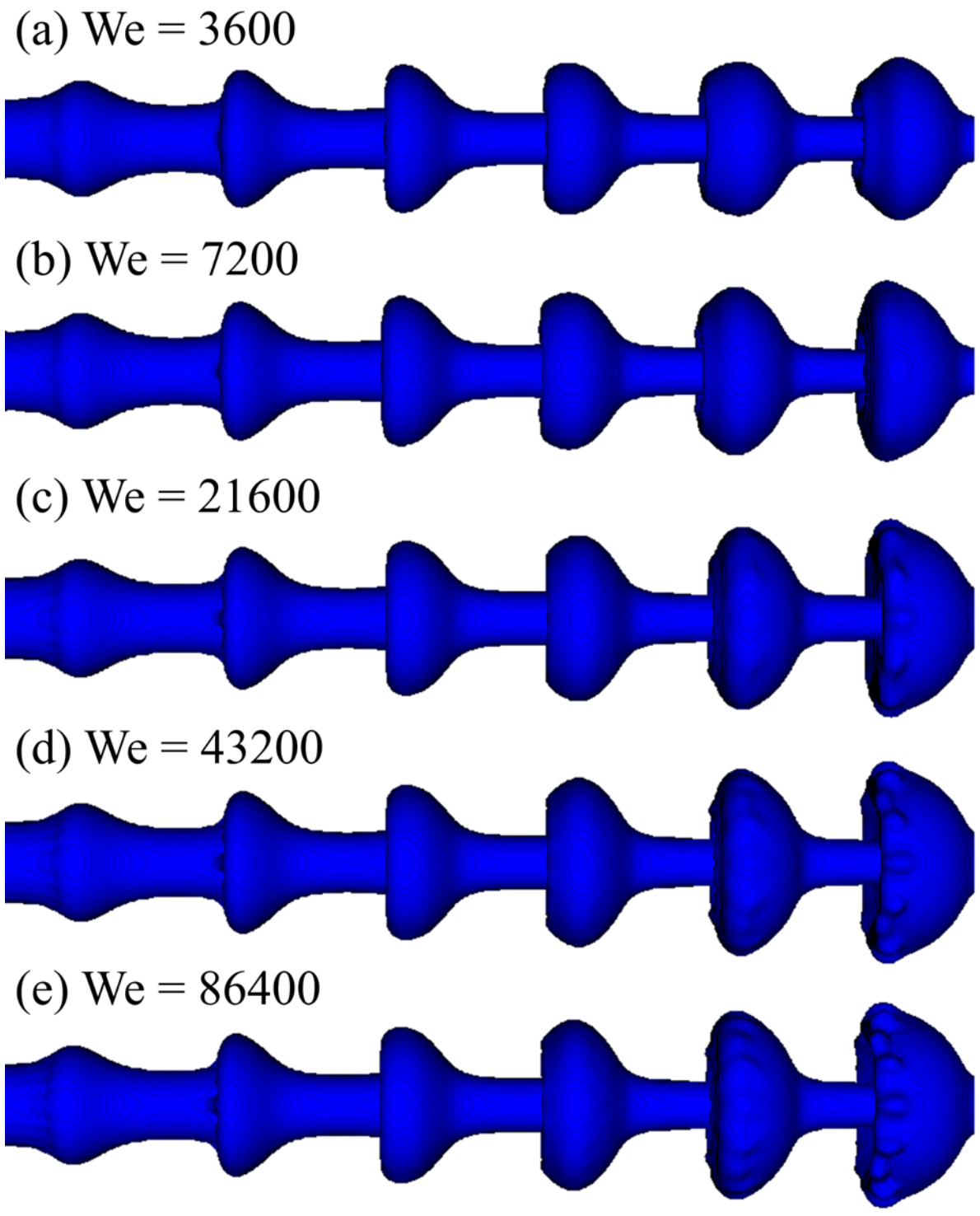

**Fig. 9. Oscillating liquid jet simulation results varying Weber number: (a) We = 3600, (b) We = 7200, (c) We = 21600, (d) We = 43200, (e) We = 86400.**

Figure 9 presents the simulation results of the oscillating liquid jet with varying surface tension coefficients. Five cases of surface tension coefficient are shown, and the Weber numbers of each case are 3600, 7200, 21600, 43200, and 86400, respectively. As the Weber number increases, the spreading angle rises due to the small surface tension. The azimuthal instability appears after the fifth oscillation. As the Weber number increases, higher azimuthal modes are observed.

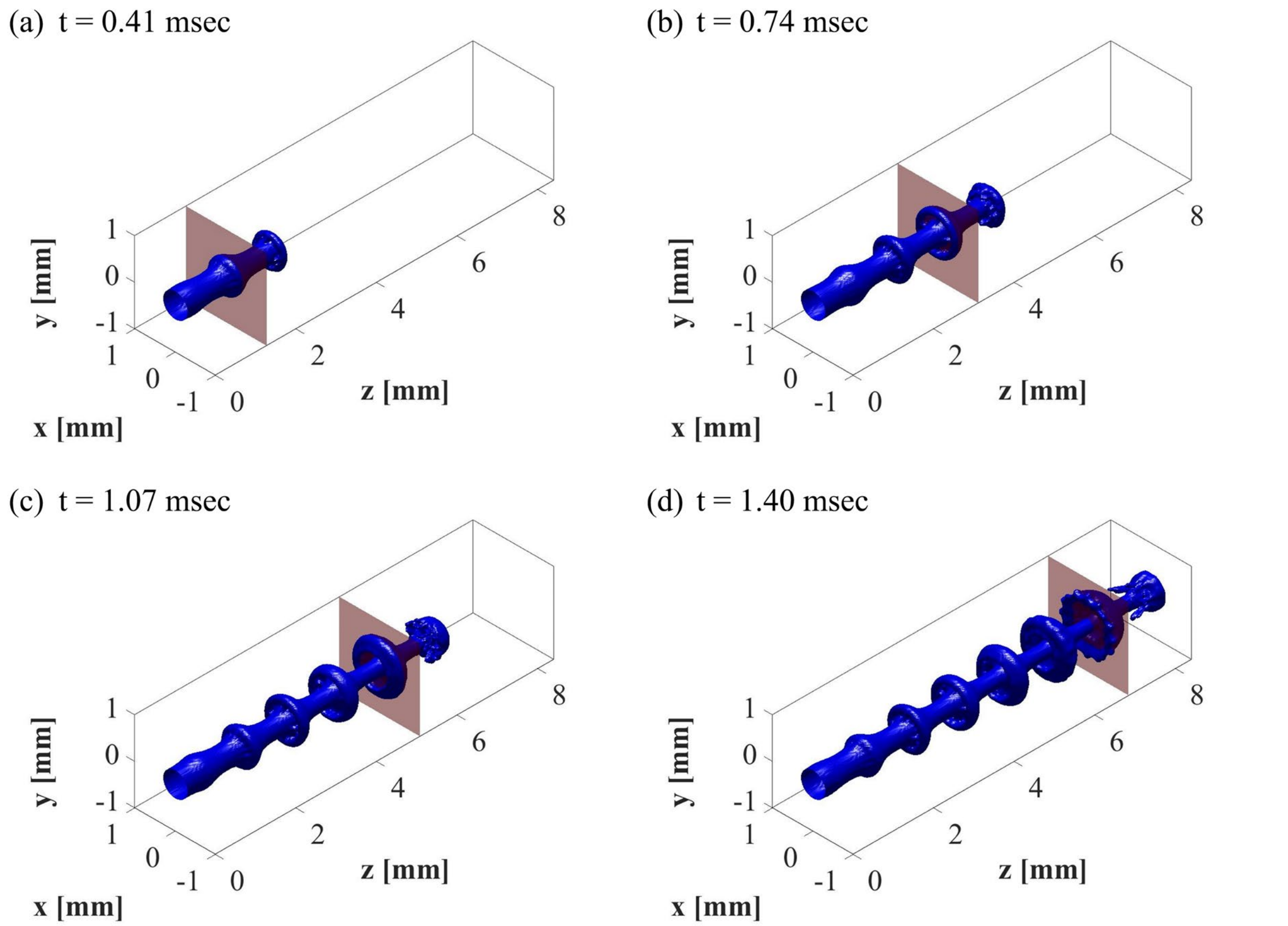

**Fig. 10. Evolution of the liquid jet shape over time using We = 86400: (a) at t = 0.408 msec, (b) at t = 0.734 msec, (c) at t = 1.071 msec, and (d) at t = 1.402 msec.**

Figure 10 depicts the evolution of the liquid jet shape over time using We = 86,400. The red plane in Fig. 10 is the plane where the radius is maximum in the tracking mushroom-shaped oscillation. As time evolves, the radius of the tracked mushroom-shaped oscillation increases. Eventually, the apparent azimuthal mode is shown at t = 1.402 msec in Fig. 10(d).

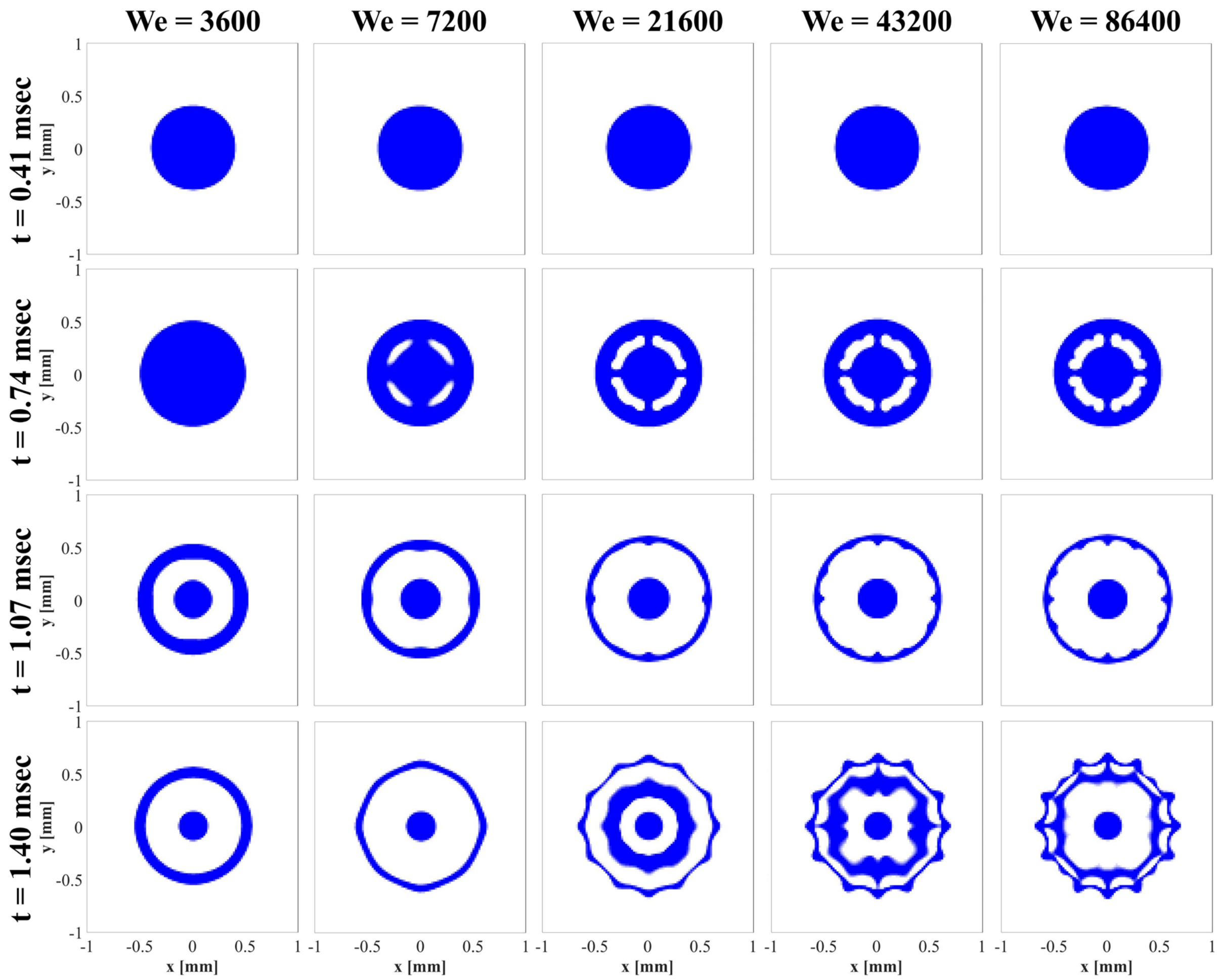

**Fig. 11. Cross-sectional view of oscillating liquid jets with various Weber numbers.**

Figure 11 shows the cross-sectional views of oscillating liquid jets with various Weber numbers. Each time step with the cross-sectional plane is equal in Fig. 10. As the Weber number increases, the radius of the liquid jet increases. Due to the vortex, the liquid jet rolls up, forming a mushroom-shaped oscillation. With increasing the Weber number, the roll-up of the liquid jet becomes more pronounced, which is captured in the cross-sectional views for We = 21600, 43200, and 86400 at t = 1.40 msec. Figure 11 also clearly shows that higher azimuthal modes are observed as the Weber number increases. The azimuthal instability initially emerges on the inner side of the mushroom-shaped oscillation (t = 1.07 msec), and subsequently develops on the outer side (t = 1.40 msec).

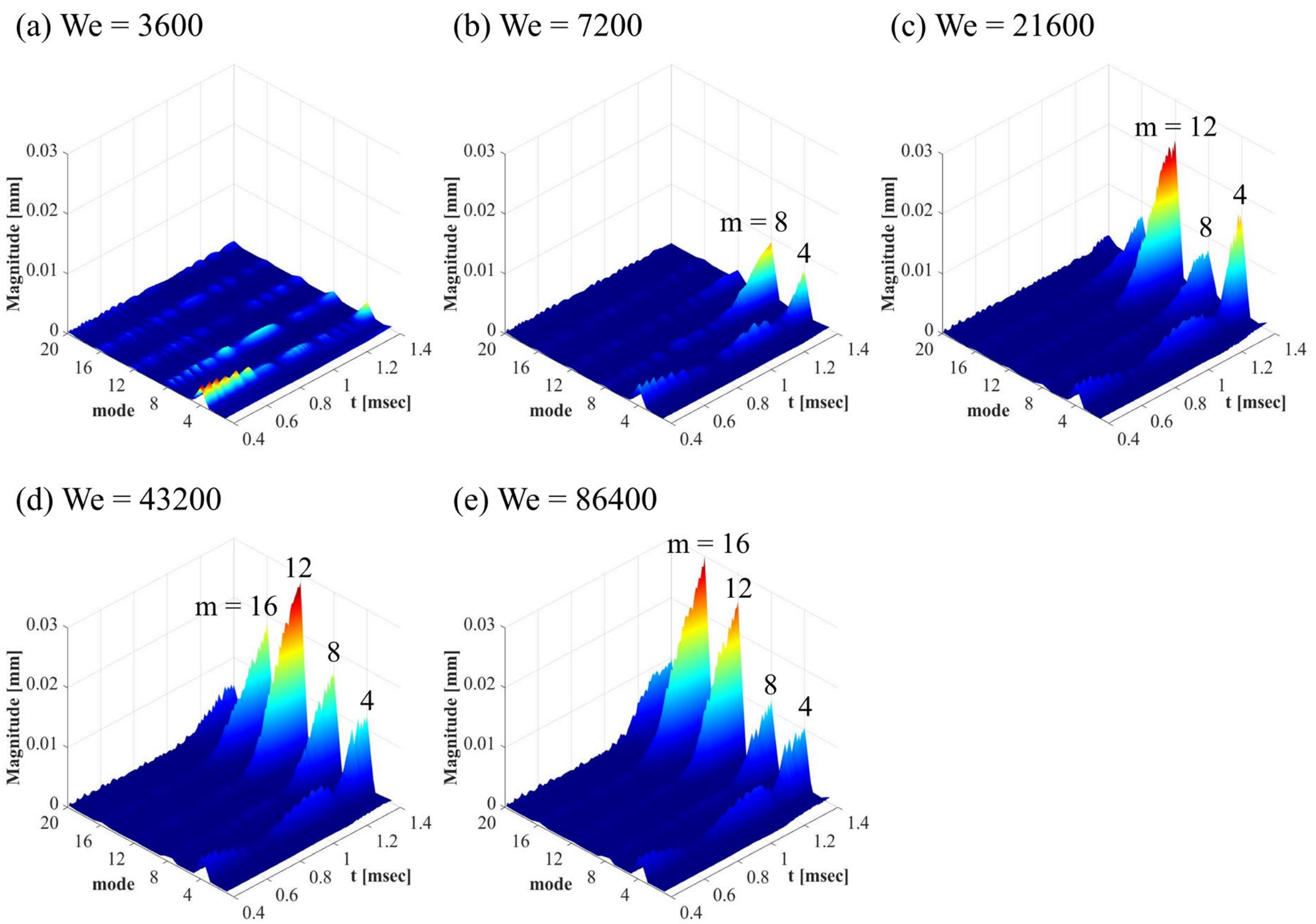

**Fig. 12. FFT magnitude results of the oscillating liquid jets: (a) We = 3600, (b) We = 7200, (c) We = 21600, (d) We = 43200, (e) We = 86400.**

Figure 12 shows the FFT results of the oscillating liquid jets following the processes of Fig. 4. The FFT analysis is conducted in the red plane in Fig. 10. In the range of Weber numbers from 3600 to 86400, modes 4, 8, 12, and 16 appear. The magnitude changes of modes 4, 8, 12, and 16 are depicted in Fig. 12. No mode appears in Fig. 12(a), where the Weber number is 3600. The magnitude of modes 4 and 8 increases after 1.2 msec, where the Weber number is 7200 in Fig. 12(b). Mode 12 is dominant in Fig. 12(c) and (d), and the magnitude increases as the Weber number increases. In Fig. 12(e), the Weber number is 86400, modes 4, 8, 12, and 16 appear after around 1 msec, and mode 16 is dominant. Thus, as the Weber number increases, the magnitude of higher azimuthal modes increases, indicating that the higher azimuthal modes are amplified.

### B. Growth rate analysis

The growth rate of azimuthal instability shows a dependence on the Weber number. As a simple approach,

the growth rate is assumed to follow a power-law dependence on the Weber number, as shown in Eq. (6).

$$\frac{A_m}{A_{ref}} = \exp\left( \mathrm{We}^n \cdot \frac{t}{T_k} \right) \tag{6}$$

where $T_k$ is constant, $A_{ref}$ is the reference amplitude at shifted time, and $n$ is the exponent of the Weber number. With rewriting the Eq. (6), the exponent of the Weber number is

$$n = \frac{\ln\left[ \frac{T_k}{t} \ln\left( \frac{A_m}{A_{ref}} \right) \right]}{\ln(\mathrm{We})}. \tag{7}$$

Using the Eq. (7), the exponent of the Weber number is chosen for modes 4, 8, 12, and 16. Figure 13 depicts the FFT results tracking one oscillation for We = 3600, 7200, 21600, 43200, and 86400 for modes 4, 8, 12, and 16. For all modes, the magnitude of the We = 3600 is small. For modes 12 and 16, the magnitude of the We = 7200 is also small. When calculating the exponent of the Weber number, We = 7200, 21600, 43200, and 86400 conditions are used for modes 4 and 8, and We = 21600, 43200, and 86400 conditions are used for modes 12 and 16. The FFT magnitudes, except for mode 4, grow after about 1msec. Mode 4 grows after about 1.2 msec. $A_{ref}$ is selected to make a constant $n$ value after 1 msec (for modes 8, 12, and 16) or 1.2 msec (for mode 4) for each Weber number condition. After making a constant $n$ value, $T_k$ is selected to make the exponent results consistent for all We number conditions. Figure 14 depicts the final results of the exponent of the Weber number for modes 4, 8, 12, and 16. The exponents of the Weber number $n$ = 0.054, 0.145, 0.3, and 0.6 satisfy all Weber number conditions simultaneously for modes 4, 8, 12, and 16, respectively. The selected $A_{ref}$ and $T_k$ values for the exponent of the Weber number are shown in Table III.

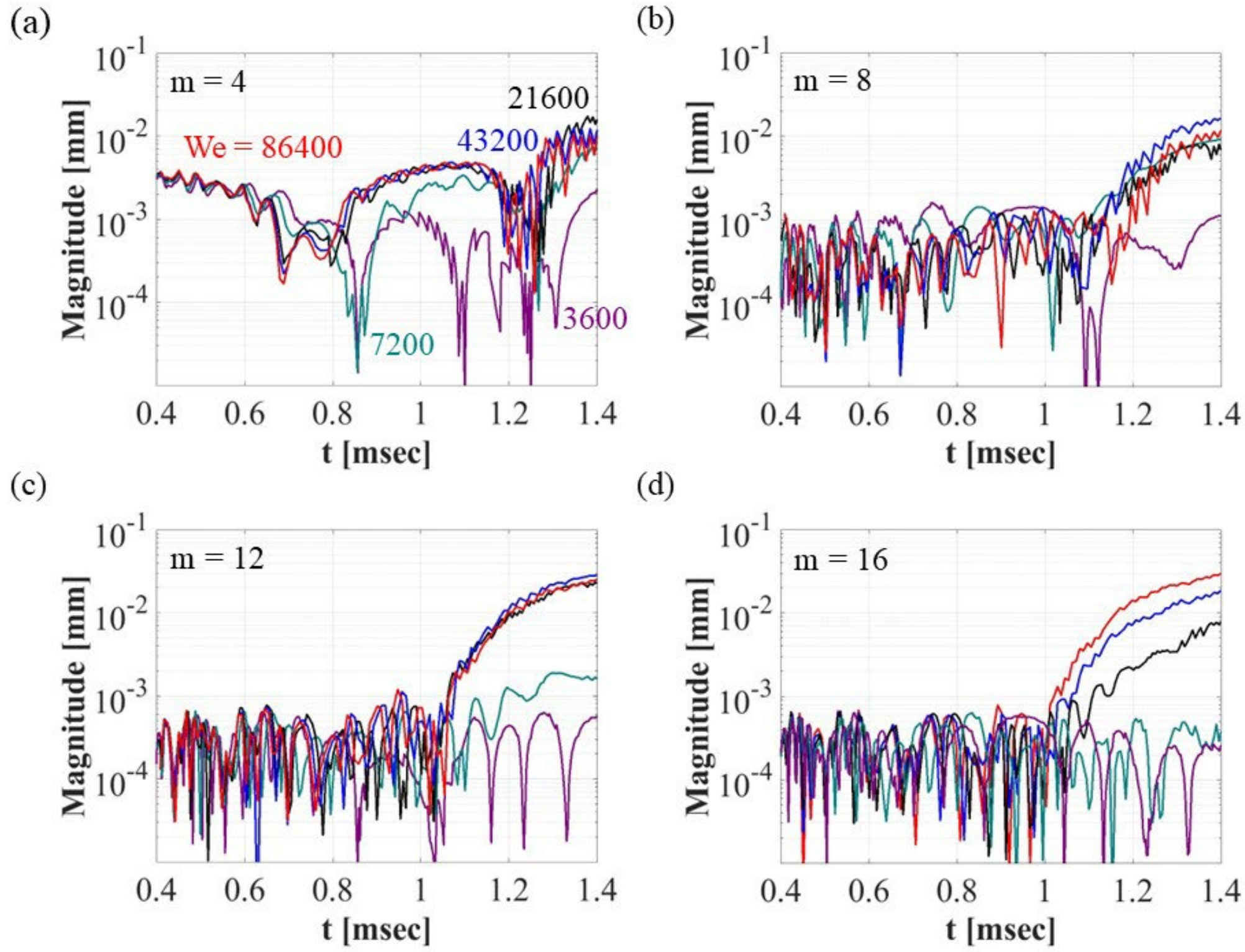

**Fig. 13. FFT results tracking one oscillation for We = 3600, 7200, 21600, 43200, and 86400 for (a) mode 4, (b) mode 8, (c) mode 12, and (d) mode 16.**

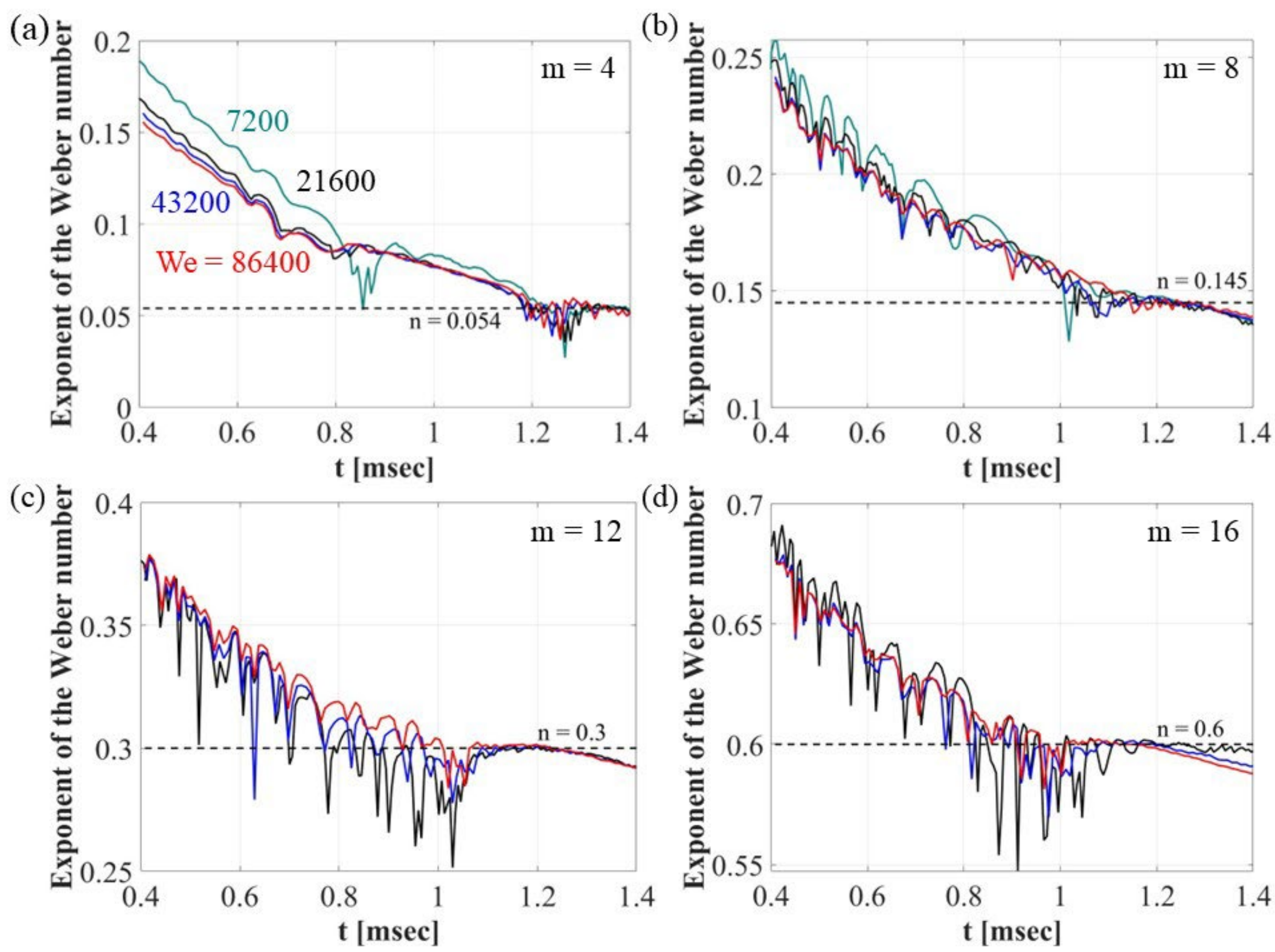

**Fig. 14. The exponent of the Weber number for (a) m = 4, (b) m = 8, (c) m = 12, and (d) m = 16.**

**Table III. The selected $A_{ref}$ and $T_k$ values for the exponent of the Weber number**

| Mode | 4 | | 8 | | 12 | | 16 | |
|---|---|---|---|---|---|---|---|---|
| We | $A_{ref}$ [mm] | $T_k$ [msec] | $A_{ref}$ [mm] | $T_k$ [msec] | $A_{ref}$ [mm] | $T_k$ [msec] | $A_{ref}$ [mm] | $T_k$ [msec] |
| 3600 | - | | - | | - | | - | |
| 7200 | $1\times10^{-9}$ | | $1.5\times10^{-10}$ | | - | | - | |
| 21600 | $1\times10^{-9}$ | 0.143 | $8\times10^{-12}$ | 0.263 | $5.5\times10^{-8}$ | 2 | $1.5\times10^{-7}$ | 50 |
| 43200 | $5\times10^{-10}$ | | $1.5\times10^{-12}$ | | $4\times10^{-9}$ | | $4\times10^{-9}$ | |
| 86400 | $2\times10^{-10}$ | | $7\times10^{-14}$ | | $1\times10^{-10}$ | | $6\times10^{-12}$ | |

Qiao et al.[21] employ linear instability analysis to study the oscillating liquid jet. The same equation that Qiao et al.[21] used in the linear instability analysis is utilized. In this work, the gas density and liquid density are 1 kg/m$^3$ and 1000 kg/m$^3$, respectively. The gas viscosity and liquid viscosity are set equal to 0.0011 kg/m·s. Thus, the dimensionless parameters to use in the linear instability analysis are shown as follows:

$$\frac{\rho_g}{\rho_l}=0.001,\ \frac{\mu_g}{\mu_l}=1,\ \mathrm{Re}=3273,\ \frac{\Delta U}{U}=0.1,\ w_s=\pi. \tag{8}$$

Here, $w_s$ is dimensionless angular frequency. The longitudinal wavelength is about 1.2 mm, which can also be calculated by dividing the velocity by the frequency. So, the longitudinal wavenumber is 0.5π.

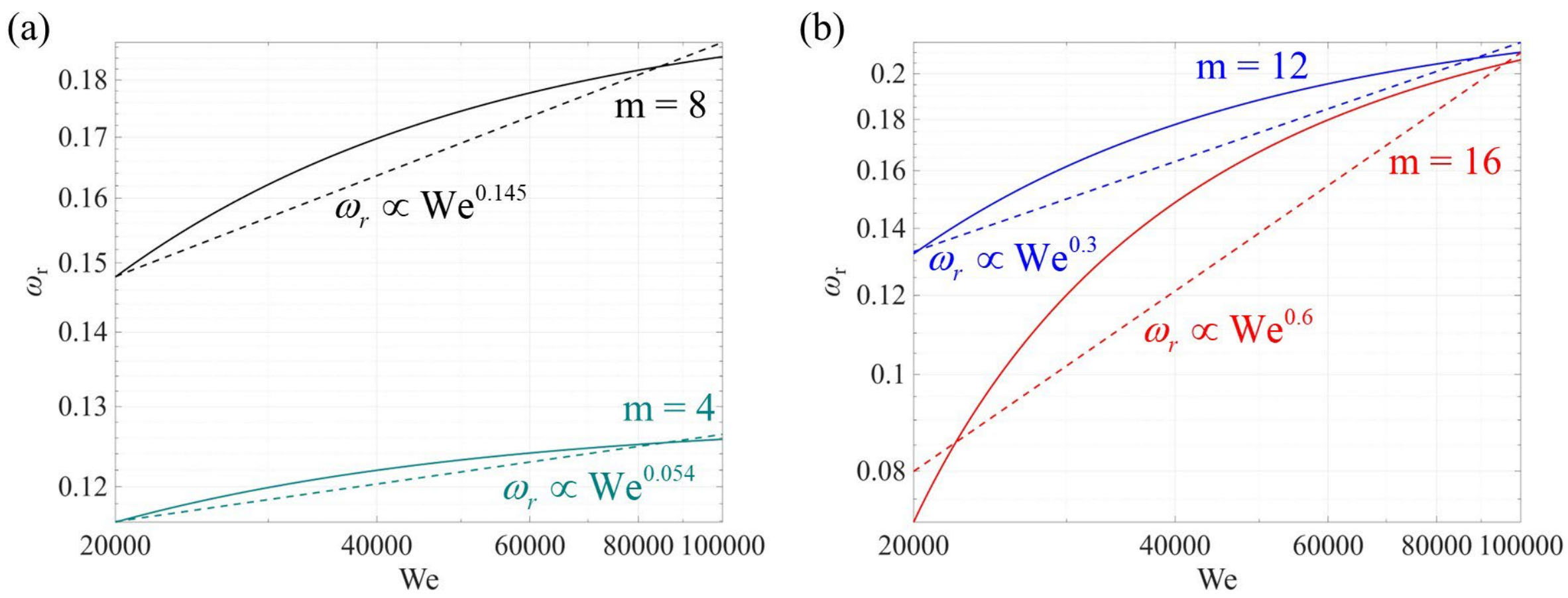

**Fig. 15. Growth rates of linear instability analysis[21] (solid lines) and estimated lines (dashed lines) for (a) m = 4, m = 8, (b) m = 12, and m = 16.**

Figure 15 depicts the growth rates obtained from the linear instability analysis[21] for Weber numbers ranging from 20,000 to 100,000, where higher azimuthal modes are observed in the simulations. The graphs are plotted on a log-log scale. In the log-log plot, the straight line represents the exponent of the Weber number, and the value obtained from Eq. (7). In this work, the exponent of the growth rate calculated from the simulation results using Eq. (7) is in good agreement with the results of the linear instability analysis for Weber numbers ranging from 20,000 to 100,000. Thus, the proposed power-law dependence of the growth rate on the Weber number provides an accurate estimate of the growth rates of oscillating liquid jets.

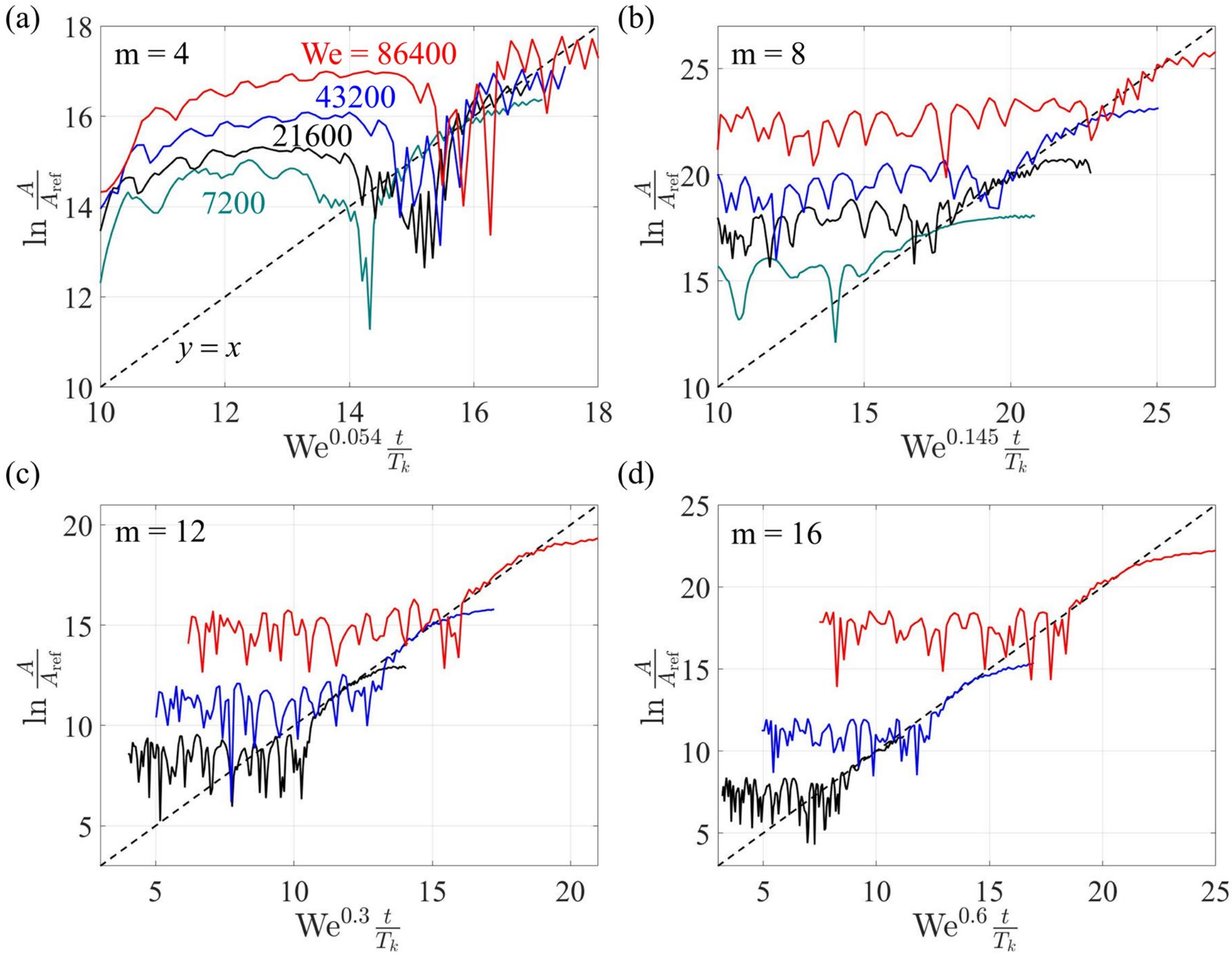

**Fig. 16. Rescaled FFT results tracking one oscillation for azimuthal modes: (a) m = 4, (b) m = 8, (c) m = 12, and (d) m = 16. The dashed line represents y = x for the reference line.**

Figure 16 shows the rescaled FFT results tracking one oscillation for azimuthal modes 4, 8, 12, and 16. The x-axis corresponds to the argument of the exponential function on the right-hand side of Eq. (6), while the y-axis represents the logarithm of the left-hand side of Eq. (6). For modes 4 and 8, the FFT results for We = 7200, 21600, 43200, and 86400 are presented. The FFT results for We = 21600, 43200, and 86400 are shown for mode 12 and 16. The magnitude of mode 16 remains small when the Weber number is 3600 and 7200 in Fig. 13(a). Thus, to clearly illustrate the shape of the other Weber numbers, the graphs for the 3600 and 7200 conditions are not plotted in Fig. 16. The dashed line is the reference line, representing $y = x$. The rescaled FFT results follow the dashed line after the onset of amplitude growth. Consistent with the comparison against the linear instability analysis, the simulation results also demonstrate that the proposed power-law dependence of the growth rate on the Weber number provides an accurate estimate of the growth rates of oscillating liquid jets.

## V. CONCLUSIONS

The oscillating liquid jet simulation is conducted using the PeleLM+IRL solver[33]. To validate the oscillating liquid jet simulation, the same dimensionless numbers, *Re*, *We*, and *St*, as the experimental result[27] are used. The mushroom-shaped oscillations appear in a similar position, and the spreading angle of the liquid jet is similar. The azimuthal mode of an oscillating liquid jet simulation is 12, reproducing azimuthal instability similar to that observed experimentally. Thus, the simulation setup of the PeleLM+IRL solver is suitable for analyzing azimuthal instability of oscillating liquid jets.

Using this simulation setup, the oscillating liquid jet simulations are performed by varying the Weber number to investigate the azimuthal instability of jets. As the Weber number increases, the spreading angle rises due to the small surface tension, and higher azimuthal modes are observed. Due to the vortex, the liquid jet rolls up, forming a mushroom-shaped oscillation. The roll-up of the liquid jet becomes more pronounced with increasing the Weber number. Fast Fourier Transform (FFT) analysis is used to find the exact azimuthal mode. As the Weber number increases, the magnitude of higher azimuthal modes increases, indicating that the higher azimuthal modes are amplified.

The growth rate is estimated as a power-law dependence on the Weber number. The exponents of the Weber number that satisfy all simulation conditions are 0.054, 0.145, 0.3, and 0.6 for m = 4, 8, 12, and 16, respectively. The exponent values are in good agreement with the results of the linear instability analysis for Weber numbers ranging from 20,000 to 100,000. When the time is rescaled using the obtained Weber number exponents, the FFT results for different Weber numbers collapse onto a single growth curve for each azimuthal mode. These results demonstrate that the Weber number governs not only the spreading and roll-up of oscillating liquid jets, but also the selection and amplification of azimuthal instability modes. In particular, the Weber number exponent of the growth rate increases with increasing azimuthal mode number, showing that the growth of higher-order modes has a stronger dependence on the Weber number. When time is rescaled using the obtained Weber number dependence, the growth curves at different Weber numbers collapse onto a single curve for each azimuthal mode, confirming that the Weber number scaling accurately describes the growth of the azimuthal modes. These findings provide a physical basis for understanding the emergence and amplification of higher-order azimuthal disturbances in oscillating liquid jets and their potential role in spray combustion instability.

## SUPPLEMENTARY MATERIAL

## ACKNOWLEDGMENTS

This work was partly supported by Korea Research Institute for defense Technology Planning and advancement(KRIT) grant funded by the Korea government (DAPA(Defence Acquisition Program Administration)) (No. KRIT-CT-22-030, Reusable Unmanned Space Vehicle Research Center, 2026) and by the Technology Innovation Program (or Industrial Strategic Technology Development Program-Shipbuilding and Offshore Industry Technology Development Program) (RS-2026-25546951, Development of a Combustor and Fuel Feed System for 100% Ammonia-Fired Gas Turbine) funded by the Ministry of Trade, Industry and Resources (MOTIR, Korea).

The authors are grateful to Mahesh Natarajan (Lawrence Berkeley National Laboratory), Robert Chiodi (Independent Scholar), and Marc Day (SINTEF Energi), who gave explanations and advice on PeleLM and IRL.

## AUTHOR DECLARATIONS

### Conflict of Interest

The authors have no conflicts to disclose.

### Author Contributions

**Hyoungwoo Kim:** Conceptualization (equal); Data Curation (lead); Formal Analysis (lead); Investigation (lead); Methodology (lead); Software (lead); Validation (lead); Visualization (lead); Writing – original draft (lead); Writing – review & editing (equal). **Seungwon Shin:** Software (supporting); Supervision (supporting); Validation (supporting); Writing – review & editing (supporting). **Hyosang Yoon:** Methodology (supporting); Visualization (supporting); Writing – review & editing (supporting). **Sanghyuen Park:** Software (supporting); Visualization (supporting); Writing – review & editing (supporting). **Dong-hyuk Shin:** Conceptualization(equal); Funding Acquisition (lead); Project Administration (lead); Resources (lead); Software (supporting); Supervision (lead); Writing – review & editing (lead).

## DATA AVAILABILITY

The data that support the findings of this study are available from the corresponding author upon reasonable request.

## REFERENCES

[1] T. C. Lieuwen, Unsteady combustor physics, (2012).

[2] K. S. Seo, C. H. Nam, Y. W. Moon, and Y. M. Han, "Liquid rocket engine development of the upper-stage for the Korea Space Launch Vehicle-II," EUCASS, 3-8 (2022).

[3] L. Crocco and S. I. Cheng, "High-frequency combustion instability in rocket motor with concentrated combustion," J. Am. Rocket Soc. 23(5), 301-313 (1953).

[4] T. Y. Toong, F. S. Richard, M. S. John, and Y. A. Griffin, "Mechanisms of combustion instability," Symp. Int. Combust. 10(1), 1301-1313 (1965).

[5] T. Lieuwen and B. T. Zinn, "The role of equivalence ratio oscillations in driving combustion instabilities in low NOx gas turbines," Symp. Int. Combust. 27(2), 1809-1816 (1998).

[6] V. S. Acharya, D. H. Shin, and T. Lieuwen, "Premixed flames excited by helical disturbances: flame wrinkling and heat release oscillations," J. Propul. Power 29(6), 1282-1291 (2013).

[7] J. O'Connor, "Understanding the role of flow dynamics in thermoacoustic combustion instability," Proc. Combust. Inst. 39(4), 4583-4610 (2023).

[8] J. Y. Lee, E. Lubarsky, and B. T. Zinn, ""Slow" active control of combustion instabilities by modification of liquid fuel spray properties," Proc. Combust. Inst. 30(2), 1757-1764 (2005).

[9] K. Sakaki, T. Funahashi, S. Nakaya, M. Tsue, R. Kanai, K. Suzuki, T. Inagawa, and T. Hiraiwa, "Longitudinal combustion instability of a pintle injector for a liquid rocket engine combustor," Combust. Flame 194, 115-127 (2018).

[10] T. Lee, J. Park, D. Han, and K. T. Kim, "The dynamics of multiple interacting swirl-stabilized flames in a lean-premixed gas turbine combustor," Proc. Combust. Inst. 37(4), 5137-5145 (2019).

[11] Y. Cheng, T. Jin, K. Luo, Z. Li, H. Wang, and J. Fan, "Large eddy simulations of spray combustion instability in an aero-engine combustor at elevated temperature and pressure," Aerosp. Sci. Technol. 108, 106329 (2021).

[12] L. Rayleigh, "On the capillary phenomena of jets," Proc. R. Soc. London. 29, 71-97 (1879).

[13] R. D. Reitz and F. V. Bracco, "Mechanism of atomization of a liquid jet," Phys. Fluids 25(10), 1730-1742 (1982).

[14] H. Q. Yang, "Asymmetric instability of a liquid jet," Phys. Fluids A 4(4), 681-689 (1992).

[15] A. Saha, J. D. Lee, S. Basu, and R. Kumar, "Breakup and coalescence characteristics of a hollow cone swirling spray," Phys. Fluids 24(12), 124103 (2012).

[16]S. Wang, S. Yang, S. Qiu, X. Li, D. L. S. Hung, and M. Xu, "Mechanism of flash boiling bubble breakup based on rim-like structure," Fuel 329, 125345 (2022).

[17]D. H. Shin, E. S. Richardson, V. Aparece-Scutariu, Y. Minamoto, and J. H. Chen, "Fluid age-based analysis of a lifted turbulent DME jet flame DNS," Proc. Combust. Inst. 37(2), 2215-2222 (2019).

[18]R. Krechetnikov and G. M. Homsy, "Crown-forming instability phenomena in the drop splash problem," J. Colloid Interface Sci. 331(2), 555-559 (2009).

[19]Y. Zhou, R. J. Williams, P. Ramaprabhu, M. Groom, B. Thornber, A. Hillier, W. Mostert, B. Rollin, S. Balachandar, P. D. Powell, A. Mahalov, and N. Attal, "Rayleigh-Taylor and Richtmyer-Meshkov instabilities: A journey through scales," Physica D 423, 132838 (2021).

[20]M. Chai, Y. Fu, S. Zheng, Z. Hong, C. Shao, K. Luo, and J. Fan, "Detailed numerical simulation of multi-scale interface-vortex interactions of liquid jet atomization in crossflow," Int. J. Multiph. Flow 161, 104390 (2023).

[21]R. Qiao, C. Zhao, Z. Ding, K. Mu, and T. Si, "Velocity modulation on the linear instability of liquid jets in ambient gas," Phys. Fluids 36(1), 012117 (2024).

[22]J. B. Keller, S. I. Rubinow, and Y. O. Tu, "Spatial instability of a jet," Phys. Fluids 16(12), 2052-2055 (1973).

[23]D. P. Busker, A. P. G. G. Lamers, and J. K. Nieuwenhuizen, "The non-linear break-up of an inviscid liquid jet using the spatial-instability method," Chem. Eng. Sci. 44(2), 377-386 (1989).

[24]A. Chauhan, C. Maldarelli, D. S. Rumschitzki, and D. T. Papageorgiou, "An experimental investigation of the convective instability of a jet," Chem. Eng. Sci. 58(11), 2421-2432 (2003).

[25]J. L. Dressler, "High-order azimuthal instabilities on a cylindrical liquid jet driven by temporal and spatial perturbations," Phys. Fluids 10(9), 2212-2227 (1998).

[26]H. Chaves, F. Obermeier, and T. Seidel, "Fundamental investigation of the disintegration of a sinusoidally forced liquid jet," ICLASS-2000, Pasadena, USA, 1018-1025 (2000).

[27]F. Geschner, F. Obermeier, and H. Chaves, "Investigation of different phenomena of the disintegration of a sinusoidally forced liquid jet," ILASS-Europe, Zurich, Switzerland, (2001).

[28]V. Srinivasan, A. J. Salazar, and K. Saito, "Modeling the disintegration of modulated liquid jets using volume-of-fluid (VOF) methodology," Appl. Math. Model. 35(8), 3710-3730 (2011).

[29]Z. Zhang and D. H. Shin, "Effect of ambient pressure oscillation on the primary breakup of cylindrical liquid jet spray," Int. J. Spray Combust. Dyn. 12, 1756827720935553 (2020).

[30]T. T. Lim and T. B. Nickels, "Instability and reconnection in the head-on collision of two vortex rings," Nature 357, 225-227 (1992).

[31]M. Cheng, J. Lou, and T. T. Lim, "Collision and reconnection of viscous elliptic vortex rings," Phys. Fluids 31(6), 067107 (2019).

[32]R. Rioboo, C. Bauthier, J. Conti, M. Voue, and J. De Coninck, "Experimental investigation of splash and crown formation during single drop impact on wetted surfaces," Exp. Fluids 35, 648-652 (2003).

[33]H. Kim, R. Chiodi, M. Day, Y. J. Kim, and D. H. Shin, "A geometric volume of fluid-based multiphase flow solver extension to the reacting flow solver, PeleLM," Adv. Eng. Softw. 213, 104074 (2026).

[34]M. S. Day and J. B. Bell, "Numerical simulation of laminar reacting flows with complex chemistry," Combust. Theory Model. 4(4), 535 (2000).

[35]A. Nonaka, J. B. Bell, M. S. Day, C. Gilet, A. S. Almgren, and M. L. Minion, "A deferred correction coupling strategy for low Mach number flow with complex chemistry," Combust. Theory Model. 16(6), 1053-1088 (2012).

[36]A. S. Almgren, J. B. Bell, P. Colella, L. H. Howell, and M. L. Welcome, "A conservative adaptive projection method for the variable density incompressible Navier-Stokes equations," J. Comput. Phys. 142(1), 1-46 (1998).

[37]B. Lafaurie, C. Nardone, R. Scardovelli, S. Zaleski, and G. Zanetti, "Modelling merging and fragmentation in multiphase flows with SURFER," J. Comput. Phys. 113(1), 134-147 (1994).

[38]M. Owkes and O. Desjardins, "A computational framework for conservative, three-dimensional, unsplit, geometric transport with application to the volume-of-fluid (VOF) method," J. Comput. Phys. 270, 587-612 (2014).

[39]R. Chiodi and O. Desjardins, "General, robust, and efficient polyhedron intersection in the Interface Reconstruction Library," J. Comput. Phys. 449, 110787 (2022).

[40]J. E. Pilliod Jr and E. G. Puckett, "Second-order accurate volume-of-fluid algorithms for tracking material interfaces," J. Comput Phys. 199(2), 465-502 (2004).